\documentclass[aps,prl,reprint,nofootinbib,twocolumn,superscriptaddress,showpacs,showkeys,longbibliography]{revtex4-1}

\usepackage{eurosym}
\usepackage{amsmath,amssymb,amstext}
\usepackage{ulem} 
\usepackage[usenames,dvipsnames]{color}
\usepackage{graphicx}
\usepackage{braket}
\usepackage{natbib}
\usepackage{comment}
\usepackage{dcolumn}
\usepackage[english]{babel}
\usepackage{wasysym}
\usepackage[colorlinks,bookmarks=false,citecolor=blue,linkcolor=red,urlcolor=blue]{hyperref}

\begin{document}
\title{Dynamical Buildup of a Quantized Hall Response from Non-Topological States} 
\author{Ying Hu}
\affiliation{Institute for Quantum Optics and Quantum Information of the Austrian Academy of Sciences, 6020 Innsbruck, Austria}

\author{Peter Zoller}
\affiliation{Institute for Quantum Optics and Quantum Information of the Austrian Academy of Sciences, 6020 Innsbruck, Austria}
\affiliation{Institute for Theoretical Physics, University of Innsbruck, 6020 Innsbruck, Austria}

\author{Jan Carl Budich}
\affiliation{Institute for Quantum Optics and Quantum Information of the Austrian Academy of Sciences, 6020 Innsbruck, Austria}
\affiliation{Institute for Theoretical Physics, University of Innsbruck, 6020 Innsbruck, Austria}


\date{\today}
\begin{abstract}
We consider a two-dimensional system initialized in a  topologically trivial state before its Hamiltonian is ramped through a phase transition into a Chern insulator regime. This scenario is motivated by current experiments with ultracold atomic gases aimed at realizing time-dependent dynamics in topological insulators. Our main findings are twofold. First, considering coherent dynamics, the non-equilibrium Hall response is found to approach a topologically quantized time averaged value in the limit of slow but non-adiabatic parameter ramps, even though the Chern number of the state remains trivial. Second, adding dephasing, the destruction of quantum coherence is found to stabilize this Hall response, while the Chern number generically becomes undefined. We provide a geometric picture of this phenomenology in terms of the time-dependent Berry curvature.
\end{abstract}

\maketitle

\begin{figure}[h!]
\centering
\includegraphics[width=0.98\columnwidth]{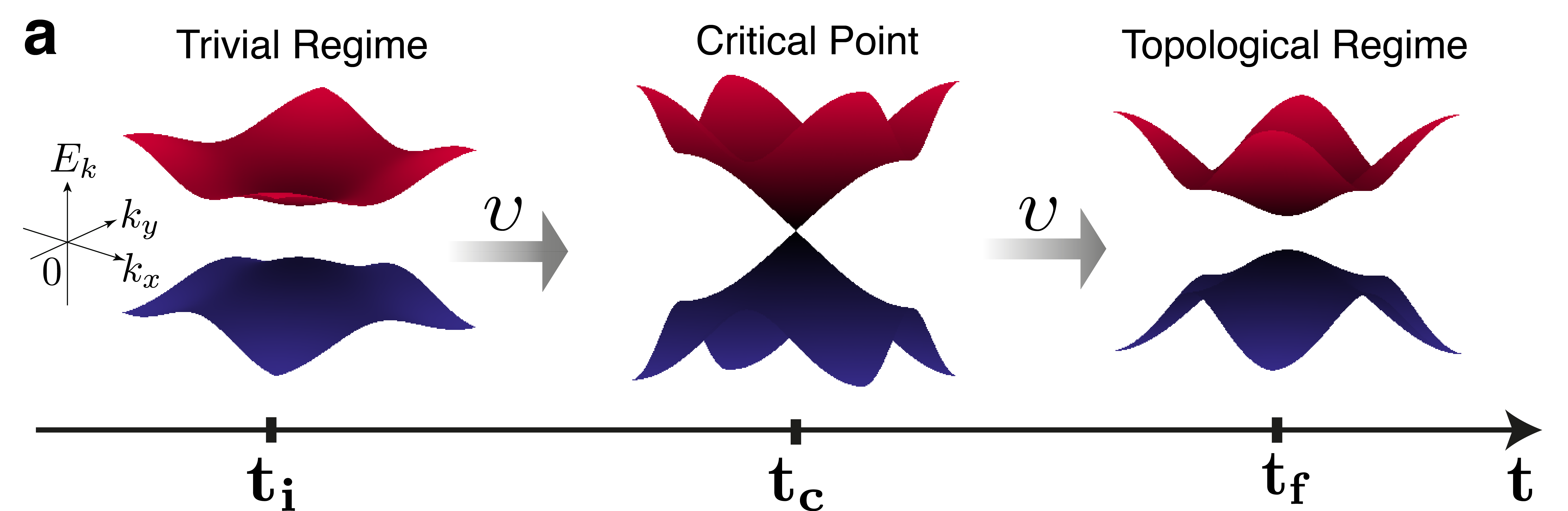}
\includegraphics[width=0.92\columnwidth]{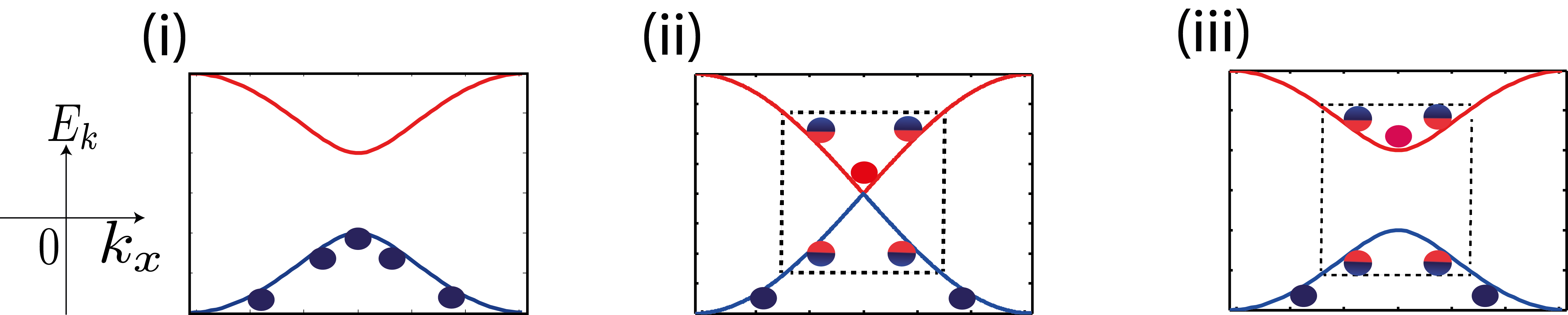}
\includegraphics[width=1\columnwidth]{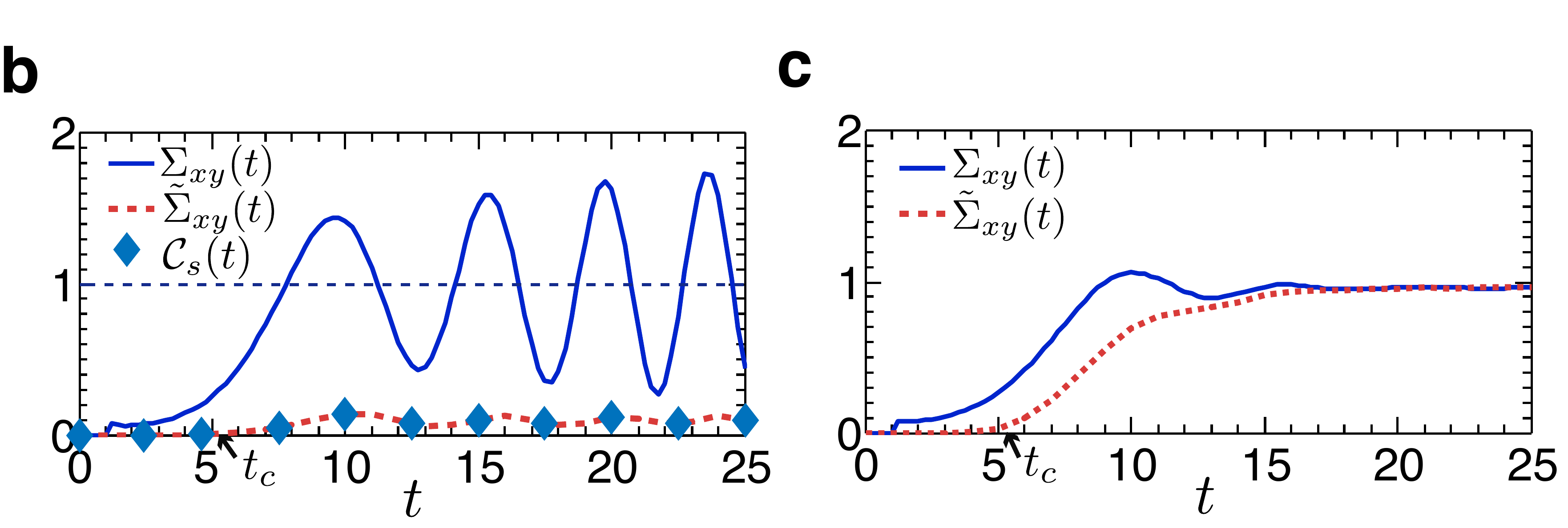}
\caption{\label{fig:one} (color online). (a) {Parameter ramp in the system Hamiltonian from non-topologial to topological regime through a phase transition. The insets show cuts of the band structure along the $k_x$ axis ($k_y=0$), illustrating (i) the initial state, (ii) the creation of excitations near the energy gap closing point, and (iii) the non-equilibrium state after the transition.} (b-c) Non-equilibrium Hall response $\Sigma_{xy}(t)$ [see Eq.~(\ref{eqn:neqhall})] and quasi-static ansatz $\tilde \Sigma_{xy}(t)$ [see Eq.~(\ref{eq:ansatz})], with (b) coherent and (c) dephasing dynamics for dephasing rate $\gamma_k=0.15$, for the ramp  $m(t)=m_i+(m_f-m_i)[1-\exp(-vt)],~m_i=-2.7, m_f=-1.0, v=0.1$ of the Hamiltonian (\ref{eqn:citoy}). In (b), $\mathcal C_s$ of the pure state [see Eq.~(\ref{eqn:chern})] trivially equals $\tilde\Sigma_{xy}$, and the dashed horizontal line denotes the long-time average of $\Sigma_{xy}$. System size $120\times120$ sites in all simulations. The finite size \cite{sup2} causes a small deviation of $\mathcal{C}_s$ from zero in (b).}
\end{figure}

{\textit{Introduction.}} Exploring the unique properties of topological insulators \cite{HasanKane2010, QiReview2011} such as Chern insulators \cite{Haldane1988} has become a major focus of research in physics. At zero temperature, the direct correspondence between the Chern number of the ground state, the Hall conductance, and the chiral edge states is well established \cite{Thouless1982,Bernevigbook}.  By contrast, far from thermal equilibrium the topological properties of the time-dependent Hamiltonian and the state may not concur \cite{Netanel2011, Gurarie2013,Foster2014,Rigol2015,CaioPreprint}, and their relation to natural observables is a subject of ongoing open discussion \cite{Kitagawa2011,Oka2009,Rudner2013, Torres2014, Hossein2015_1,Wang2015,Jan2016}. Yet, such non-equilibrium scenarios generically occur in present experiments on ultracold gases \cite{Aidelsburger2011,Struck2012,Atala2014,Aidelsburger2013,Miyake2013,Jotzu2014,Aidelsburger2015,Goldman2015,Flaschner2015}, where starting from a topologically trivial initial state, the Hamiltonian is driven into a topological parameter regime, thus going through a topological quantum phase transition [cf. Fig. \ref{fig:one} (a)]. However, the Chern number of the state is well known to remain zero under coherent dynamics. This topological discrepancy between the actual state vs.  the Hamiltonian immediately raises the challenge as to which manifestations of topology can be observed, even without entering a Chern insulator state, i.e. without adiabatically following the ground state. Below, we report two major theoretical contributions to address this issue, which may also shed light on ongoing experiments aimed at observing quantum Hall physics with cold atoms.

\begin{figure}[h!]
\centering
\includegraphics[width=1.01\columnwidth]{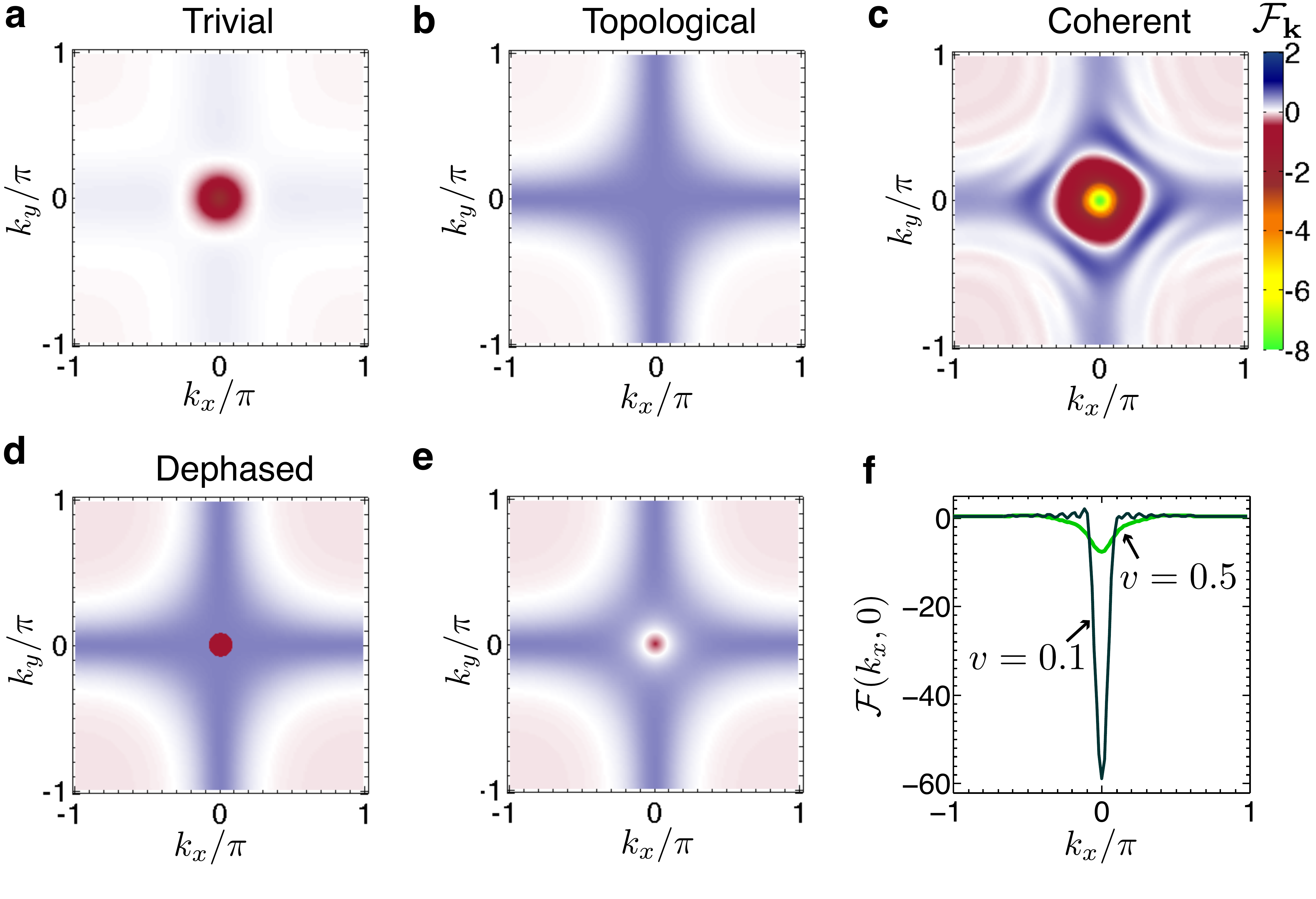}
\caption{\label{fig:two} (color online). Berry curvature $\mathcal F_k$. (a)-(b): Lower band $\mathcal F_k$ for (a) initial Hamiltonian $H(m_i)$ and (b) final Hamiltonian $H(m_f)$. (c): $\mathcal F_k(t)$ for coherently evolved state at $t>t_c$. (d): Discountinous $\mathcal F_k$ for the dephased steady state. Corresponding weighted curvature $\sqrt{p}_k\mathcal F_k$ (integrand of Eq.~(\ref{eq:ansatz}) is shown in panel (e). (f): $\mathcal F(k_x,0)$ of the coherently evolved states at $t>t_c$,  for $v=0.1$ and $v=0.5$, respectively. The simulations are done with a local adaptive method in momentum space resolving system sizes of up to $1795\times 1795$ sites. Ramp velocity $v=0.5$ (c)-(e), $m_i=-2.5,m_f=-1$.}
\end{figure}

\begin{figure}[h!]
\centering
\includegraphics[width=0.83\columnwidth]{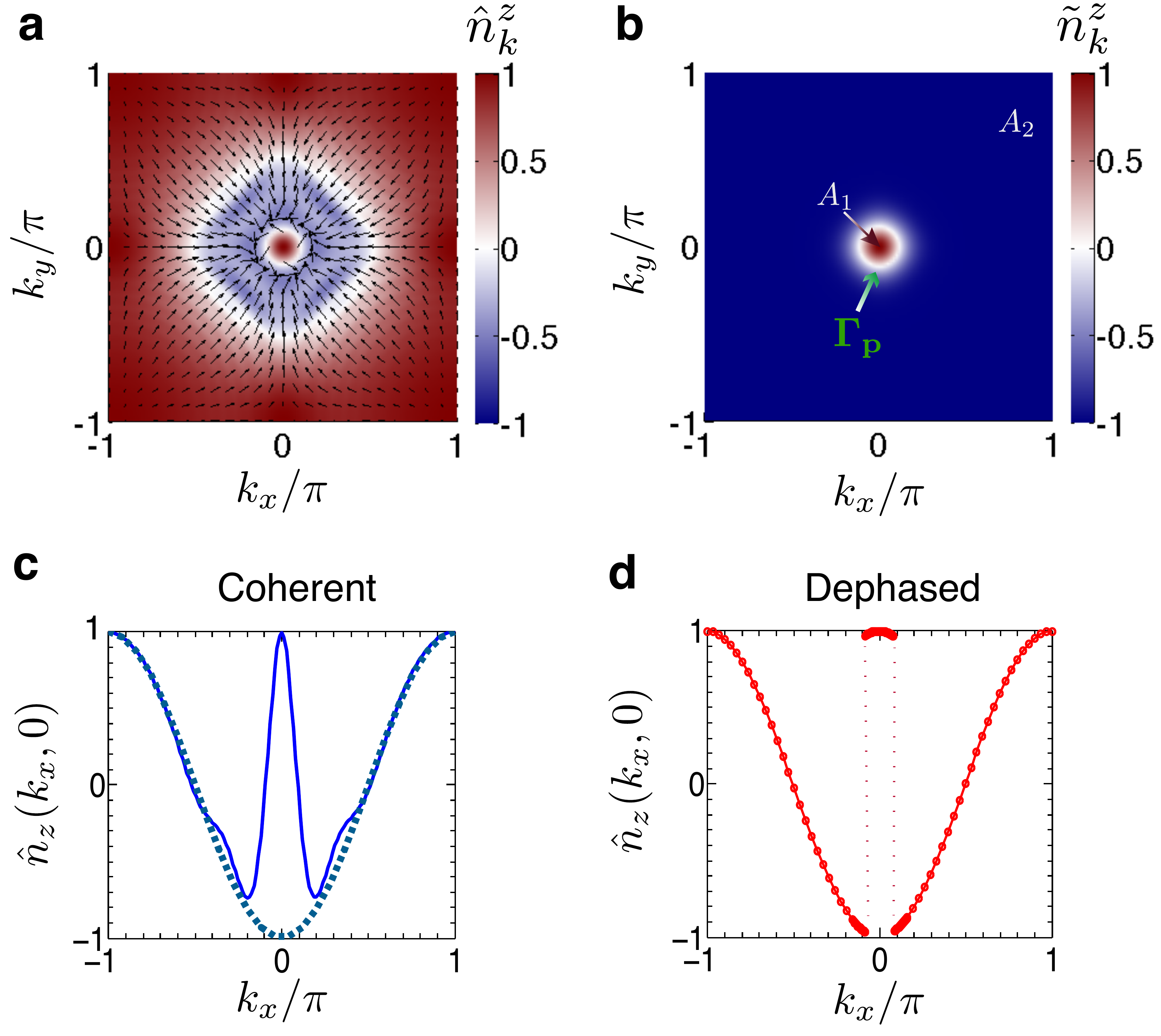}
\caption{\label{fig:three} (color online). (a): Bloch vector $\hat{n}_k$ of the coherently evolving state at $t\gg t_c$. Arrows depict in-plane configuration $(\hat{n}_k^x,\hat{n}_k^y)$, whereas $n_k^z$ is indicated with color. (b): Occupation in the eigenbasis of $H(m_f)$ parameterized by $\tilde{n}_k^z$. Contour $\Gamma_p$ (closed white curve) defined by $\tilde n_k^z=0$, i.e. at equal weight superposition of the upper and lower band. (c): $\hat{n}_z(k_x,0)$ as a smooth function of $k_x$ for $k_y=0$ for the coherently time-evolved state (blue solid) and ground state of $H(m_f)$ (blue dashed). (d): $\hat{n}_k^z(k_x,0)$ for the dephased steady state, which exhibits a discontinuous jump around purity gap closing point. $m_i=-2.5$, $m_f=-1$, $v=0.5$ in all plots. $\gamma_k=0.5$ in (d). System size is $120\times 120$ sites in all plots. }
\end{figure}

First, we show that the non-equilibrium bulk Hall response can be \textit{quantized} -- at least in an asymptotic sense -- reflecting uniquely the topology of the \textit{instantaneous} Hamiltonian, despite the non-topological nature of the state at all times. Our main result on the coherent dynamics is shown in Fig.~\ref{fig:one} (b): A non-equilibrium Hall response exhibiting strongly oscillatory behavior in time is found to build up when the Hamiltonian enters a Chern insulator regime. The time-averaged Hall response at large times approaches a topologically quantized value in the limit of slow but non-adiabatic ramps. Second, we find that adding classical noise induced \textit{dephasing} not only stabilizes this Hall response [see Fig.~\ref{fig:one} (c)], but also allows for a geometric interpretation that eludes the standard notion of Chern numbers in closed systems [see Figs.~\ref{fig:two} and \ref{fig:three}]. The central entity underlying this picture is the time-dependent Berry curvature of the density matrix describing the mixed state of the open system. In particular, we find that the Berry curvature generically acquires \textit{discontinuities} [see Fig. \ref{fig:two} (d) and Fig. \ref{fig:three} (d)] that render the Chern number not well-defined. This is found to originate from the interplay of the Landau-Zener (LZ) dynamics \cite{LandauZener} around the gap closing and dephasing without energy relaxation. These observations allow us to explain the behavior of the Hall response, including its dependence on the ramp velocity and asymptotic quantization. While recent studies have focused on the coherent dynamics of chiral edge states \cite{Rigol2015,CaioPreprint}, our present theoretical findings reveal a conclusive picture of bulk response properties in non-equilibrium Chern insulators beyond the coherent framework.

{{\textit{Topological discrepancy: Hamiltonian vs. state.}}} As a paradigmatic example \cite{footGeneral} of a Chern insulator \cite{Haldane1988} exhibiting quantum Hall physics, we consider a time-dependent lattice version of the massive 2D Dirac Hamiltonian \cite{Qi2008}
\begin{align}
\!\!\!H(m(t))=\sum_k c_k^\dag H_k(m(t)) c_k=\sum_k c_k^\dag\left[\vec d_k(m(t))\cdot \vec \sigma\right] c_k.
\label{eqn:citoy}
\end{align}
Here, $c_k$ denotes the two-spinor of fermionic operators at lattice momentum $k$, and $\vec d_k(m(t))=\left(\sin(k_x),\sin(k_y),m(t)+\cos(k_x)+\cos(k_y)\right)$, where energy is measured in units of the hopping strength. For fixed $m$, the lower band of $H_k(m)$ has Chern number $\mathcal C=-\text{sgn}(m)$ for $0<\lvert m \rvert<2$, while $\mathcal C=0$ otherwise. In the following, we will focus on the experimental relevant situation where the topology of $H(m(t))$ changes from trivial to non-trivial as $m(t)=m_i + (m_f-m_i)[1-\text{e}^{-vt}]$ $(t\ge 0)$ is ramped from $m_i$ to $m_f$ with velocity $v$, undergoing a topological transition with an energy gap closing at time $t=t_c$ and momentum $k_c=0$ (see Fig. \ref{fig:one} (a)). The initial state is assumed to be the insulating ground state of Hamiltonian $H(m_i)$ at half filling, i.e. a topologically trivial state. 

To account for the generically mixed states appearing in the open system dynamics, we consider the time-dependent density matrix $\rho(t)$. Assuming the conservation of lattice-translation invariance, $\rho(t)$ factorizes into the components  $\rho_k(t)=\frac{1}{2}[1+\vec n_k(t)\cdot \vec \sigma]$ at lattice momentum $k$ in the first Brillouin zone (BZ), where $\vec \sigma$ denotes the standard Pauli matrices. The vector $\vec n_k$ describes the polarization of $\rho_k$ on the Bloch sphere and its length $p_k=\lvert \vec n_k\rvert^2\le 1$ measures the purity of the state which has been coined purity gap \cite{DiehlTopDiss,BardynTopDiss,DissCI,TopDens}. 
For $p_k(t)>0$, topologically inequivalent states at time $t$ are distinguished by the instantaneous Chern number \cite{Chern}
\begin{align}
\label{eqn:chern}
&\mathcal C_s(t) \quad=\quad \frac{1}{2\pi}\int_{\text{BZ}}d^2{k}\mathcal F_k(t),
\end{align} 
where the Berry curvature is defined as
\begin{align}
\label{eqn:curvature}
\mathcal F_k=-\frac{1}{2}\hat n_k\cdot \left [\left(\partial_{k_x}\hat n_k\right)\times\left(\partial_{k_y}\hat n_k\right)\right]
\end{align}
with $\hat n_k=\vec n_k/\sqrt{p_k}$. For $p_k\equiv 1$, $\mathcal C_s$ reduces to the standard Chern number of a pure state. 

Under coherent evolution which simply acts as a smooth unitary transformation on $\rho_k(t)$, $\mathcal C_s$ is constant in time. Here, while the Hamiltonian (\ref{eqn:citoy}) enters a topologically non-trivial Chern insulator regime for $t>t_c$, the Chern number of the state $\mathcal C_s\equiv 0$ at all times. Beyond coherent dynamics where the state generically becomes mixed with $p_k<1$, $\mathcal C_s$ is protected by the purity gap provided it is finite. If the purity gap closes, i.e. $p_k=0$ for some $k$, $C_s$ becomes undefined.

{{\textit{Non-Equilibrium bulk Hall response.}}} We dynamically probe the non-equilibrium Hall response 
\begin{align}
\Sigma_{xy}(t)=\frac{1}{E_x}\int_{\textrm{BZ}}d^2{k}\textrm{Tr}[{j}_y\rho_k(t)],
\label{eqn:neqhall}
\end{align}
where the current $j_y$ in $y$-direction is generated by a small electric field $E_x$ in $x$-direction (the 2D system is defined in the $xy$-plane), and we measure conductance in units of ${e^2}/{h}$. To probe the Hall response of the system, we switch on a \textit{small} homogenous electric field at $t=0$ as $E_x(t)=E_0(1-\exp(-t/\tau_e))$ as generated by a spatially homogeneous time-dependent vector potential, thus preserving translation invariance. In experiments on cold atoms in optical lattices, such an electric field can be synthetically generated \cite{Aidelsburger2015}. In our simulations, we choose $\tau_e=5.0$ and $E_0=0.001$, so that the electric field is sufficiently small to have negligible effect on the state.

To gain intuition for the non-equilibrium nature of Hall response $\Sigma_{xy}(t)$, we formally interpret $\rho_k(t)$ at every point in time as a canonical thermal density matrix associated with a (fictitious) Hamiltonian $\tilde H_k$, i.e., $\rho_k\sim \text{e}^{-\tilde H_k}$. In this picture, the corresponding equilibrium Hall conductance $\tilde \Sigma_{xy}(t)$ as derived \cite{sup} from the standard Kubo formula \cite{Mahan} reads as
\begin{align}
\tilde \Sigma_{xy}(t)=\frac{1}{2\pi}\int_{\textrm{BZ}} \text{d}^2 k\sqrt{p_k(t)}\mathcal{F}_k(t).
\label{eq:ansatz}
\end{align}
The deviation of $\tilde{\Sigma}_{xy}(t)$ from the exact value $\Sigma_{xy}(t)$ serves as a measure of how different the non-equilibrium Hall response is from its equilibrium counterpart associated with the same instantaneous state. Note that even if the purity gap $p_k$ closes, $\tilde \Sigma_{xy}$ stays well defined by the definition $\sqrt{p_k} \mathcal{F}_k=0$ for $\sqrt{p_k}=0$. The intuition behind this continuation is that $\sqrt{p_k}=0$ represents an infinite temperature state which does not contribute to the Hall conductance.

{{\textit{Quantized Hall response without Chern insulator state.}}} As a first main result, we show the non-equilibrium Hall response under coherent dynamics where the Chern number $C_s(t)$ is pinned to zero at all times [see Fig. \ref{fig:one} (b)]. During the \textit{non-adiabatic} ramp of $m(t)$ through the gap closing [see Fig. \ref{fig:one} (a)], the non-equilibrium population of the eigenstates of $H(m(t))$ is determined by LZ physics \cite{LandauZener}: Away from $k_c$ where the energy gap is larger than the ramp velocity $v$ at all times, the system stays in the ground state, while close to $k_c$ excitations and coherent superpositions of ground and excited states, respectively, are created. Right at $k_c$, the excited state is populated with probability one, thus rendering $\rho(t)$ \textit{orthogonal} to the ground state of the final Hamiltonian $H(m_f)$. We note that for pure states, $\tilde \Sigma_{xy}=\mathcal C_s$. Hence, also $\tilde \Sigma_{xy}$ has to stay zero at all times in the thermodynamic limit. By contrast, a significant non-equilibrium Hall response $\Sigma_{xy}(t)$ which shows a strongly oscillatory behavior is found to build up dynamically [see Fig. \ref{fig:one} (b)]. In even stronger disagreement with the zero Chern number, its time averaged value over many oscillation periods approaches asymptotically the quantized value of a Chern band in the limit of small $v$. Our subsequent analysis regarding the influence of dephasing will give a geometrical picture reconciling this discrepancy.

{{\textit{Stablization of Hall response by dephasing.}}} We now show that adding classical noise to the dynamics, which induces dephasing, the oscillations of the Hall response $\Sigma_{xy}$, as shown in 
Fig. \ref{fig:one} (b) for coherent evolution, damp out. This yields the {\it smooth Hall response} plotted in Fig. \ref{fig:one} (c) -- well captured by $\tilde \Sigma_{xy}$ for $t\gg t_c$ --  which becomes {\it quantized in the slow ramp limit}. In the noisy dephasing dynamics  the coherent superpositions of excited and ground states of $H_k(m(t))$ are randomized, as described by the master equation for the stochastically averaged density matrix \cite{Lindblad1976,Gardiner2010book,Ying2015} 
\begin{align}
\dot \rho_k=-i [H_k(m),\rho_k]+\gamma_k\left[\tilde \sigma_k^z\rho_k\tilde \sigma_k^z- \rho_k\right].
\label{eqn:LindbladDeph}
\end{align}  
Here $\tilde \sigma_k^i(t)$ denote the standard Pauli-matrices in the basis of the instantaneous Hamiltonian $H_k(m(t))$ at lattice momentum $k$. In addition to the Hamiltonian part, {{Eq.~(\ref{eqn:LindbladDeph}) contains a noise-induced pure dephasing term, which preserves the population of the instantaneous eigenstates of $H_k(m(t))$ and thus the average energy, whilst the relative phase coherence decays at a rate $\gamma_k$ \cite{footnote1}.}}  The time-evolution generated by Eq.~(\ref{eqn:LindbladDeph}) does not preserve the purity of the averaged density matrix.


Such a dephasing appears naturally in cold atom experiments with natural or engineered laser noise, where system parameters become stochastic functions of time. We note that laser fluctuations act as a temporal {\it global} noise, which uniformly affects the system.  In particular, a fluctuating (global) mass parameter can result from frequency fluctuations or modulation of the laser light, while (global) fluctuations in the hopping amplitude can arise from intensity variations. The dephasing terms in the Hamiltonian underlying  (\ref{eqn:LindbladDeph}) are assumed to commute with the system Hamiltonian, as discussed in Ref.~ \cite{Hannes2012}. In the limit of fast fluctuations (white noise), 
the stochastically averaged density matrix averaged obeys the master equation (\ref{eqn:LindbladDeph}).

While the values of both $\Sigma_{xy}$ and $\tilde \Sigma_{xy}$ approach the quantized value reflecting the Chern insulator Hamiltonian in the slow ramp limit $v\ll1$ \cite{foot2}, the Chern number of the steady state generically becomes un-defined due to a purity gap closing. To gain a deeper understanding of this phenomenon, we  study below the time dependent geometric properties of the state described by the stochastically averaged density operator.

{{\textit{Discontinuous Berry curvature and geometric analysis.}}} In Fig. \ref{fig:two}, we compare the Berry curvature [see Eq. (\ref{eqn:curvature})] of the dephased and the coherently evolved states to that of the ground states of both $H(m_i)$ and $H(m_f)$. Remarkably, for the dephased state, $\mathcal F_k$ exhibits characteristic discontinuities [see Fig. \ref{fig:two} (d)] that will allow us to explain the behavior of the Hall response reported above. Note that the experimental observation of the Berry curvature for a system of ultracold atoms has recently been reported \cite{Flaschner2015}.

To reveal the effects of dephasing, we represent the density matrix in the eigenbasis of the instantaneous Hamiltonian $H(m(t))$, here denoted by $\rho_k(t)=\frac{1}{2}(1+\tilde n_k(t)\cdot \tilde \sigma_k(t))$. In this basis, the occupation of the upper band at momentum $k$ is simply given by $(1+\tilde n_k^z)/2$, and dephased stead state is of the form\begin{align}
\rho_k^s=\frac{1}{2}(1+\tilde n_k^z \tilde \sigma_k^z)=\frac{1}{2}(1+ \tilde n_k^z \hat d_k(m_f)\cdot \vec\sigma_k)
\label{eqn:perfectdeph}
\end{align}
that is diagonal in the basis of the final Hamiltonian $H(m_f)$ with $\hat{d}_k=\vec{d}_k/|\vec{d}_k|$ and has purity $\lvert\tilde n_k^z\rvert^2$. 
From the coherent LZ dynamics at $t>t_c$, we expect $\tilde n_k^z\approx 1$ close to the gap closing momentum $k_c$ and $\tilde n_k^z\approx -1$ far away from $k_c$  [see Fig.~\ref{fig:three} (b)]. 
Hence, there must be a closed contour $\Gamma_p$ around $k_c$ in the BZ for $t>t_c$, where the pure system state is an equal weight superposition of the lower and the upper band ($\tilde n_k^z(t)=0$). On $\Gamma_p$, dephasing results in a completely mixed steady state $\rho_k^s =\frac{1}{2}$, implying a purity gap closing in the long time limit. To visualize this behavior, we show the Berry curvature [see Fig. \ref{fig:two}] and the Bloch sphere vector $\hat n_k$ of the density matrix $\rho_k$ [see Fig. \ref{fig:three}]. In the coherent case $\hat n_k^z$  stays {\textit{smooth}} [see Fig.~\ref{fig:three} (a),(c)], even though  with decreasing $v$ the change of $\hat n_k^z$ becomes more and more steep. This gives rise to a sharp peak in the Berry curvature [see Fig. \ref{fig:two} (c),(f)] which renders $\mathcal C_s$ and $\tilde \Sigma_{xy}$ zero, irrespective of $v$. By contrast, this peak in $\mathcal F_k$ is absent in the dephased steady state [see Fig. \ref{fig:two} (d),(e)]. Instead, from Eq.~(\ref{eqn:perfectdeph}), we find that
\begin{equation}
\hat{n}_k^s=\frac{\tilde n_k^z}{|\tilde n_k^z|}\hat{d}_k(m_f)=\textrm{sgn}(\tilde n_k^z)\hat{d}_k(m_f) \label{eqn:nks}
\end{equation}
which exhibits a {\textit{discontinuous}} jump by $2 \lvert \hat d_k^z(m_f)\rvert$ on $\Gamma_p$ [see Fig.~\ref{fig:three} (d)], where $\tilde n_k^z$ changes sign. This renders the mixed state Chern number $\mathcal C_s$ [see Eq.~(\ref{eqn:chern})] undefined, as the Berry curvature is not well-defined on $\Gamma_p$. However, this discontinuity does not contribute to $\tilde \Sigma_{xy}$ as it concurs with the purity gap closing $p_k=0$. As we see from the asymptotic agreement of the blue and red curves in Fig.~\ref{fig:one} (c), $\tilde \Sigma_{xy}$ provides a good intuition for the real Hall response $\Sigma_{xy}$ long after $t_c$.

To compute $\tilde \Sigma_{xy}$ [see Eq. (\ref{eq:ansatz})], the BZ is decomposed into two patches $A_1$ and $A_2$ separated by $\Gamma_p$ [see Fig. \ref{fig:three} (b)]. From Eq.~(\ref{eqn:perfectdeph}), we immediately conclude that $\mathcal F_k$ is simply the upper band (lower band) Berry curvature of the final Hamiltonian $H(m_f)$ on $A_1$ ($A_2$). The radius of $A_1$ is proportional to $v$. Therefore, in the limit of small $v$, the value of the integral over the BZ is dominated by $A_2$ and we find to leading order in $v$ \cite{sup}
\begin{equation}
\tilde\Sigma_{xy} = \mathcal C-\frac{1}{2\pi}\frac{v}{|m_f+2|} \label{eqn:Sigma5}
\end{equation}
approaching the value corresponding to the Chern number $\mathcal C=1$ of the lower band of $H(m_f)$. 
This reconciles the behavior of the Hall response with the underlying mixed state geometry, contrasting the discrepancy between $\mathcal C_s$ and $\Sigma_{xy}$ in the coherent dynamics.

{{\textit{Concluding discussion.}}} Our present analysis has been based on translation-invariant systems of free fermions. However, our key results are found to be robust in the presence of various imperfections that may occur in real experimental settings. In particular, we have carefully verified that both a trapping potential and weak static disorder only lead to minor quantitative changes in the Hall response \cite{sup}. Regarding many-body interactions, the nearly insulating character of the state is expected to limit the influence of multi particle scattering on the bulk response properties.

In summary, we have shown how the topology of the instantaneous Hamiltonian can manifest itself in the bulk response of a system far from thermal equilibrium, even if its state stays non-topological. In the presence of dephasing we were able to provide a geometric explanation of this phenomenon which goes beyond the well-established framework of topological quantum numbers in closed systems. These results are of immediate relevance for current experiments on synthetic material systems where the preparation of topologically non-trivial Hamiltonians is state of the art while prepairing their ground state, or at least a low temperature thermal state, remains an open challenge.

{\textit{Note added. }} After the submission of this manuscript two preprints focusing on the non-equilibrium Hall response in the coherent quench dynamics starting from a topologically non-trivial initial state have appeared on the arXiv \cite{RefaelPre,CooperPre}.

{\textit{Acknowledgment.}}  We acknowledge discussions with M. Baranov, N. Goldman, H. Jiang, and H. Pichler. This project was supported by the ERC Synergy Grant UQUAM and the SFB FoQuS (FWF Project No. F4016-N23). Y. H. also acknowledges the support from the Institut f\"{u}r Quanteninformation GmbH.

\bibliographystyle{apsrev}

\begin{thebibliography}{14}
\expandafter\ifx\csname natexlab\endcsname\relax\def\natexlab#1{#1}\fi
\expandafter\ifx\csname bibnamefont\endcsname\relax
\expandafter\ifx\csname bibfnamefont\endcsname\relax
  \def\bibfnamefont#1{#1}\fi
\expandafter\ifx\csname citenamefont\endcsname\relax
  \def\citenamefont#1{#1}\fi
\expandafter\ifx\csname url\endcsname\relax
  \def\url#1{\texttt{#1}}\fi
\expandafter\ifx\csname urlprefix\endcsname\relax\def\urlprefix{URL }\fi
\providecommand{\bibinfo}[2]{#2}
\providecommand{\eprint}[2][]{\url{#2}}


\bibitem{HasanKane2010}{M. Z. Hasan and C. L. Kane, Rev. Mod. Phys. {\bf{82}}, 3045 (2010).}

\bibitem{QiReview2011}{X. L. Qi and S. C. Zhang, Rev. Mod. Phys.  {\bf{83}}, 1057 (2011).}

\bibitem{Haldane1988}{F. D. M. Haldane, Phys. Rev. Lett. {\bf 61}, 2015 (1988).}

\bibitem{Thouless1982}{D. J. Thouless, M. Kohmoto, M. P. Nightingale, and M. den Nijs, Phys. Rev. Lett. {\bf 49}, 405 (1982).}

\bibitem{Bernevigbook} {B. A. Bernevig and T. L. Hughes, \textit{Topological Insulators and Topological Superconductors} (Princeton University Press, Princeton, NJ, 2013).}



\bibitem{Netanel2011}{N. H. Lindner, G. Refael, and V. Galitski, Nat. Phys.  {\bf 7}, 490 (2011).}

\bibitem{Gurarie2013}{M. S. Foster, M. Dzero, V. Gurarie, and E. A. Yuzbashyan, Phys. Rev. B {\bf 88}, 104511 (2013).}

\bibitem{Foster2014}{M. S. Foster, V. Gurarie, M. Dzero, and E. A. Yuzbashyan, Phys. Rev. Lett. {\bf 113}, 076403 (2014).}

\bibitem{Rigol2015}{L. D'Alessio and M. Rigol, Nat. Commun. {\bf 6}, 8336 (2015).}
\bibitem{CaioPreprint}{M. D. Caio, N. R. Cooper, and M. J. Bhaseen, Phys. Rev. Lett. {\bf{115}}, 236403 (2015).}

\bibitem{Jan2016}{J. C. Budich and M. Heyl, Phys. Rev. B {\bf 93}, 085416 (2016).}


\bibitem{Kitagawa2011}{T. Kitagawa, T. Oka, A. Brataas, L. Fu, and E. Demler,  Phys. Rev. B {\bf 84}, 235108 (2011).}

\bibitem{Oka2009}{T. Oka and H. Aoki, Phys. Rev. B {\bf 79}, 081406 (R) (2009).}

\bibitem{Rudner2013}{M. S. Rudner, N. H. Lindner, E. Berg, and M. Levin, Phys. Rev. X {\bf 3}, 031005 (2013).}

\bibitem{Torres2014}{L. E. F. Foa Torres, P. M. Perez-Piskunow, C. A. Balseiro, and G. Usaj, Phys. Rev. Lett. {\bf 113}, 266801 (2014). }


\bibitem{Hossein2015_1}{H. Dehghani, T. Oka, and A. Mitra, Phys. Rev. B {\bf 91}, 155422  (2015).}

\bibitem{Wang2015}{P. Wang, M. Schmitt, and S. Kehrein, Phys. Rev. B {\bf 93}, 085134  (2016).}





\bibitem{Aidelsburger2011}{M. Aidelsburger, M. Atala, S. Nascimb\'ene, S. Trotzky, Y. A. Chen, and I. Bloch, Phys. Rev. Lett. {\bf 107},
255301 (2011).}   

\bibitem{Struck2012}{J. Struck, C. \"{O}lschl\"{a}ger, M. Weinberg, P. Hauke, J. Simonet, A. Eckardt, M. Lewenstein, K. Sengstock, and P. Windpassinger, Phys. Rev. Lett. {\bf 108},
225304 (2012).}

\bibitem{Atala2014}{M. Atala, M. Aidelsburger, M. Lohse, J. T. Barreiro, B. Paredes, and I. Bloch, Nat. Phys. {\bf 10}, 588 (2014).}
\bibitem{Aidelsburger2013}{M. Aidelsburger, M. Atala, M. Lohse, J. T. Barreiro, B. Paredes, and I. Bloch, Phys. Rev. Lett. {\bf 111},
185301 (2013).}
\bibitem{Miyake2013}{H. Miyake, G. A. Siviloglou, C. J. Kennedy, W. C. Burton, and W. Ketterle, Phys. Rev. Lett. {\bf 111},
185302 (2013).}
\bibitem{Jotzu2014}{G. Jotzu, M. Messer, R. Desbuquois, M. Lebrat,
T. Uehlinger, D. Greif, and T. Esslinger, Nature (London) {\bf 515},
237 (2014).}
\bibitem{Aidelsburger2015}{M. Aidelsburger, M. Lohse, C. Schweizer, M. Atala, J. T. Barreiro, S. Nascimb\'ene, N. R. Cooper, I. Bloch, and N. Goldman, Nat. Phys. {\bf 11}, 162 (2015).}
\bibitem{Goldman2015}{N. Goldman, N. Cooper,  and J. Dalibard, arXiv: 1507.07805 (2015).}
\bibitem{Flaschner2015}{N. Fl\"{a}schner, B. S. Rem, M. Tarnowski, D. Vogel, D. -S. L\"{u}hmann, K. Sengstock, C. Weitenberg, arXiv: 1509.05763 (2015).}

\bibitem{LandauZener} {L. D. Landau and E. M. Lifshitz, \textit{Quantum Mechanics} (Pergamon, New York, 1958); C. Zener, \textrm{Proc. R. Soc. A}
\textbf{137}, 696 (1932).}

\bibitem{footGeneral}{
We note that our analysis can readily be generalized beyond this minimal setting of two-banded models.}

\bibitem{Qi2008}{X. L. Qi, T. L. Hughes, and S. C. Zhang, Phys. Rev. B {\bf{78}}, 195424 (2008).} 

\bibitem{sup2}{Regarding the relevance of finite sized effects in our simulations, see also the supplemental material.}

\bibitem{DiehlTopDiss}{
S. Diehl, E. Rico, M. Baranov, and P. Zoller,
Nat. Phys. {\bf{7}}, 971 (2011).}

\bibitem{BardynTopDiss} {C. E. Bardyn, M. A. Baranov, C. V. Kraus, E. Rico, A. Imamoglu, P. Zoller, and S. Diehl, New J. Phys. {\bf 15}, 085001  (2013).}

\bibitem{DissCI}{
J. C. Budich, P. Zoller, and S. Diehl, Phys. Rev. A {\bf{91}}, 042117 (2015).}

\bibitem{TopDens}{
J. C. Budich and S. Diehl, Phys. Rev. B {\bf{91}}, 165140 (2015).}


\bibitem{Chern}{
S. S. Chern, Ann. Math. {\bf{47}}, 85 (1946).}

\bibitem{sup}{For details, see also the supplemental material.}

\bibitem{Mahan}
G. D. Mahan, {\textit{Many-Particle Physics}} (Springer 2000).

\bibitem{Lindblad1976}{G. Lindblad, Commun. Math. Phys. {\bf 48}, 119 (1976).}

\bibitem{Gardiner2010book} {C. Gardiner, \textit{Stochastic Methods: A
Handbook for the Natural and Social Sciences} (Springer Berlin Heidelberg,
2010).}

\bibitem{Ying2015}
Y. Hu, Z. Cai, M. A. Baranov, and P. Zoller, Phys. Rev. B {\bf{92}}, 165118 (2015).  




\bibitem{footnote1}{We have carefully checked that all results reported below are qualitatively unchanged if dephasing is switched on only
after the gap closing time $t_c$.}

\bibitem{Hannes2012}{H. Pichler, J. Schachenmayer, J. Simon, P. Zoller, and A. J. Daley, Phys. Rev. A {\bf 86}, 051605 (R) (2012).}

\bibitem{foot2}{We note that extremely strong dephasing, i.e. $\gamma_k$ larger than the energy scale of $H_k(m_f)$ has a detrimental influence on the current thus suppressing the Hall response $\Sigma_{xy}$. In the experimentally natural regime $\gamma_k < 1$, this effect is not visible and can only be observed if very strong dephasing can be engineered.}

\bibitem{RefaelPre}
J. H. Wilson, J. C. W. Song, and G. Refael, arXiv:1603.01621 (2016).

\bibitem{CooperPre}
M. D. Caio, N. R. Cooper, and M. J. Bahseen, arXiv:1604.04216 (2016).




  
\end{thebibliography}

\end{document}


\title{Supplemental Material for: \\ Dynamical Buildup of a Quantized Hall Response from Non-Topological States}

\author{Ying Hu}
\affiliation{Institute for Quantum Optics and Quantum Information of the Austrian Academy of Sciences, 6020 Innsbruck, Austria}

\author{Peter Zoller}
\affiliation{Institute for Quantum Optics and Quantum Information of the Austrian Academy of Sciences, 6020 Innsbruck, Austria}
\affiliation{Institute for Theoretical Physics, University of Innsbruck, 6020 Innsbruck, Austria}

\author{Jan Carl Budich}
\affiliation{Institute for Quantum Optics and Quantum Information of the Austrian Academy of Sciences, 6020 Innsbruck, Austria}
\affiliation{Institute for Theoretical Physics, University of Innsbruck, 6020 Innsbruck, Austria}

\date{\today}

\maketitle

\section{Derivation of $\tilde \Sigma_{xy}$}

In this section we derive the analytical form (5) of $\tilde\Sigma_{xy}(t)$ in the main text. The key idea behind $\tilde\Sigma_{xy}$ is to formally interpret the time-dependent state $\rho(t)$ as a \emph{static} thermal equilibrium ensemble at every point in time, so that a {\emph{fictitious}} equilibrium Hall response can be computed by means of the standard Kubo formula.

Consider the density matrix $\rho_k(t)=\frac{1}{2}[1+\vec{n}_k(t)\cdot\vec{\sigma}]$ at lattice momentum $k$. Using the definition of the purity $p_k=|\vec{n}_k|^2$, we have the spectral decomposition
\begin{equation}
\rho_k(t)=\frac{1+\sqrt{p_k(t)}}{2}|+\rangle\langle+|+\frac{1-\sqrt{p_k(t)}}{2}|-\rangle\langle-|, \label{eqn:spectral}
\end{equation}
where $|\pm\rangle$ denote the eigenstates of $\rho_k$ which satisfy 
\begin{equation}
\vec{n}_k\cdot \vec{\sigma} |\pm\rangle=\pm\sqrt{p_k}|\pm\rangle. \label{eqn:eigen}
\end{equation}
Formally, $\rho_k$ may be interpreted as a thermal state of a fictitious Hamiltonian $\tilde{H_k}$ at unit temperature at every lattice momentum, i.e.
\begin{equation}
\rho_k=\frac{1}{Z}\text{e}^{-\tilde H_k}.
\end{equation}
We note that $\tilde H_k$ has the same eigenstates $\lvert \pm\rangle$ as $\rho_k$. From Eq. (\ref{eqn:eigen}) we then see that the Berry curvature of the two bands of $\tilde H_k$ is given by $\mathcal F_k^\pm=\pm\hat{n}_k\cdot (\partial_{k_x}\hat{n}_k\times\partial_{k_y}\hat{n}_k)=\pm \mathcal{F}_k$. From the standard Kubo formula in this equilibrium picture \cite{Mahan}, we obtain a Hall response which is given by the integral of Berry the curvature weighted by the occupation probability in the upper (lower) band of $\tilde H_k$ (see Eq. (\ref{eqn:spectral})), i.e., 
\begin{eqnarray}
\tilde\Sigma_{xy}(t)&=&\int \frac{d{\bf k}}{2\pi} \left[\mathcal{F}_k^+(t)\frac{1+\sqrt{p_k(t)}}{2}+\mathcal{F}_k^-(t)\frac{1-\sqrt{p_k(t)}}{2}\right] \nonumber\\
&=&\int d{\bf k}\sqrt{p_k}\mathcal{F}_k.  \label{Sigma}
\end{eqnarray}
Thus we arrive at Eq. (5) in the main text. 

\section{$\tilde \Sigma_{xy}$ for dephased steady states}\label{Sup_Sec3}

In this section, we provide details for the evaluation of $\tilde\Sigma_{xy}$  (see Eq. (5) in the main text) for the steady state of a system after a slow passage of the Hamiltonian (see Eq. (1) in the main text) from a topologically trivial to a  nontrivial regime in the presence of dephasing. Concretely, the mass parameter $m(t)=m_i+(m_f-m_i)(1-\text{e}^{-vt})$ is ramped from $m_i<-2$ to $m_f$ (with $-2< m_f<0$) with a ramp velocity $v$. In this case, the energy gap of the system closes at lattice momenta $k_c=(0,0)$ at the critical time $t_c=-\log(\frac{m_f+2}{m_f-m_i})/v$ defined by $m(t_c)=-2$. 

\subsection{Exact analysis} 

We start from Eq. (7) in the main text for the steady state density matrix, $\rho_k^s=(1/2)[1+\tilde{n}_k^z\hat{d}^f_k\cdot\vec{\sigma}_k]$, which is diagonal in the energy eigen-basis of $H(m_f)$ assuming complete decoherence between the eigenstates. Here $\hat{d}^f_k=\vec{d}^f_k/|\vec{d}^f_k|$ with $\vec{d}^f_k=(\sin k_x,\sin k_y, m_f+\cos k_x+\cos k_y)$. For $\tilde{n}_k^z\neq 0$, the direction $\hat{n}_k^s$ of $\rho_k^s$ on the Bloch sphere reads as
\begin{equation}
\hat{n}_k^s=\frac{\tilde n_k^z}{|\tilde n_k^z|}\hat{d}^f_k=\textrm{sgn}(\tilde n_k^z)\hat{d}^f_k.\label{eqn:nks}
\end{equation}
Eq. (\ref{eqn:nks}) exhibits a jump at the contour $\Gamma_p$ where $\tilde n^z_k=0$. As discussed in the main text, $\Gamma_p$ separates the Brillouin zone (BZ) into two disconnected patches $A_1$ and $A_2$ [see Fig. 3 (b) in the main text] that additively contribute to $\tilde\Sigma_{xy}$. We can hence decompose $\tilde\Sigma_{xy}=\tilde\Sigma_{A_1}+\tilde\Sigma_{A_2}$ with
\begin{eqnarray}
\tilde\Sigma_{A_1}&=&-\frac{1}{2\pi}\int_{A_1} d^2 k |\tilde{n}^z_k| \mathcal F_k^f, \label{eqn:A1}\\
\tilde\Sigma_{A_2}&=&\frac{1}{2\pi}\int_{A_2}d^2k |\tilde{n}^z_k| \mathcal F_k^f. \label{eqn:A2}
\end{eqnarray}
where $\mathcal F_k^f=\frac{1}{2}\hat{d}^f_k\cdot \left[\left(\partial_{k_x}\hat{d}^f_k\right)\times \left(\partial_{k_y}\hat{d}^f_k\right)\right]$ denotes the Berry curvature associated with the lower band of $H(m_f)$.  Computing $\tilde\Sigma_{xy}$ using Eqs. (\ref{eqn:A1}) and (\ref{eqn:A2}) requires the knowledge of $\tilde n_k^z$ which parameterizes the occupation difference between the upper and lower energy levels of $H(m_f)$. Using that the occupation is not affected by dephasing in the eigen-basis of $H(m(t))$, we can infer $\tilde n_k^z$ from the coherent Landau-Zener dynamics of the system in a numerically exact fashion, or, as presented next, by analytical estimation in the limit of small $v$.

\begin{figure}[h!]
\centering
\includegraphics[width=0.65\columnwidth]{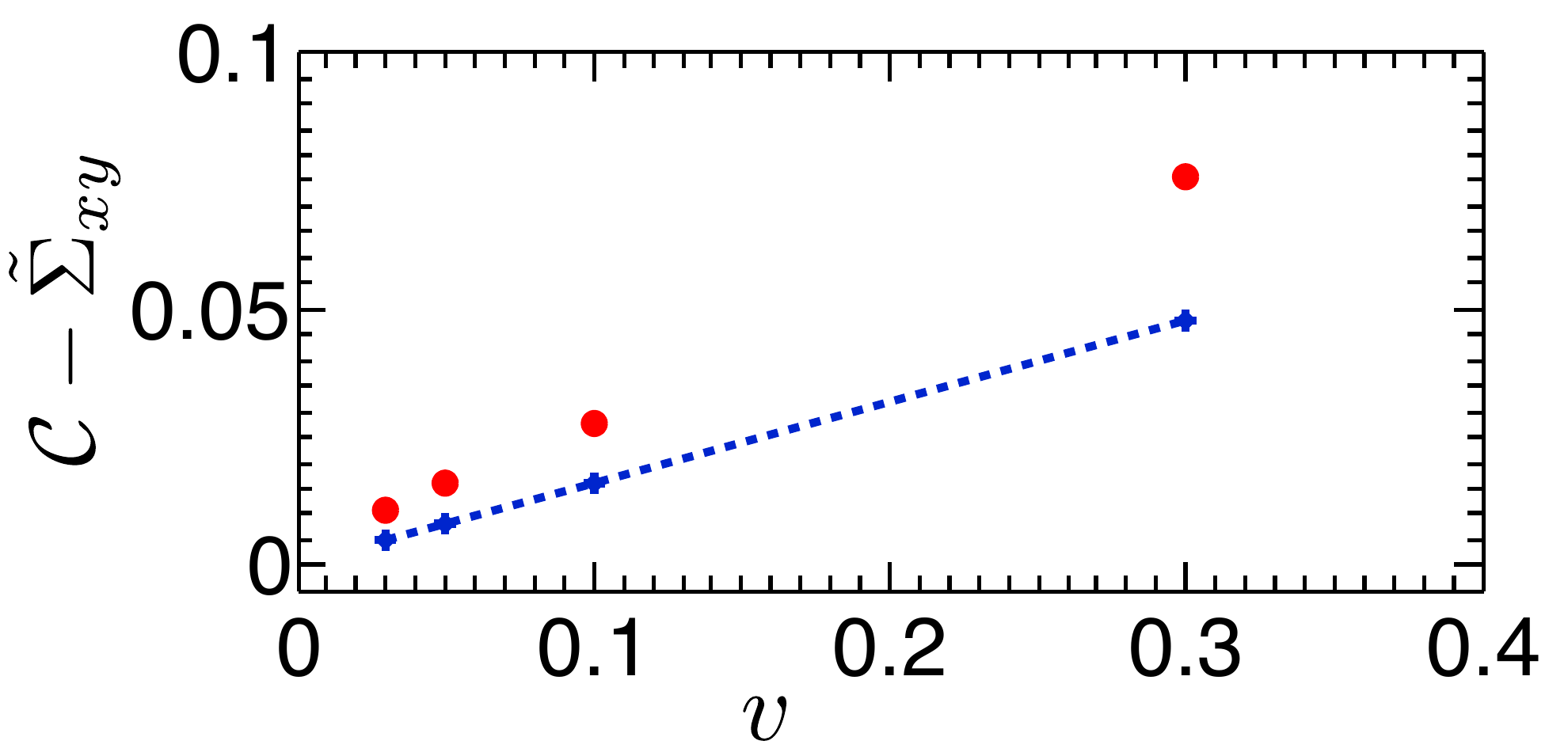}
\caption{\label{fig:one} (color online). $\mathcal C-\tilde\Sigma_{xy}$ as a function of $v$ for the dephased steady state. The numerical results (red dots) are compared with the corresponding prediction (blue curve) using Eq. (\ref{eqn:Sigma5}). For dephasing dynamics, we have taken $\gamma_k=0.15$, $m_i=-2.5$, $m_f=-1$ and $v=0.03,0.05, 0.1,0.3$, respectively. System size is $120\times 120$ sites. }\label{AppFig:1}
\end{figure}

\subsection{Analytical estimate of $\tilde \Sigma_{xy}$ for slow ramp}

If $v$ is sufficiently small, the excitations occur in a very narrow region near the gap-closing point $k_c=(0,0)$ in the BZ. There, we linearize $H_k(m(t))=c_k^\dag [\vec{d}_{k}(m(t))\cdot\vec \sigma]c_k$ around $k_c$ which yields
\begin{equation}
H_k(m(t))= k_x \sigma_x+k_y\sigma_y+[m(t)+2]\sigma_z+\mathcal O(k^2). \label{eqn:dkc}
\end{equation}
In the context of Landau-Zener dynamics \cite{LandauZener}, Eq. (\ref{eqn:dkc}) may be seen as two time-dependent unperturbed energy levels $\pm E(t) = \pm(m(t)+2)$ that are coupled by the momentum dependent terms, giving rise to a minimum separation $\Delta_k^{\mathrm{min}}\approx 2k$ at time $t_c$. 
Linearizing the time-dependence of $m(t)$ around $t_c$, the Landau-Zener velocity is given by 
\begin{align}
v_{\textrm{LZ}}=\lvert\dot m(t_c)\rvert=\lvert 2+m_f \rvert v.
\end{align}
Within these approximations, using the Landau-Zener formula \cite{LandauZener} for the probability $P_{\textrm{LZ}}^k$ of exciting the two-level system at momentum $k$ during a linear ramp through the avoided level crossing gives
 \begin{equation}
\tilde n_k^z = 2P_{\textrm{LZ}}^k-1= 2e^{-\frac{\pi k^2}{v_{\textrm{LZ}}}}-1. \label{eqn:lz}
\end{equation}
It immediately follows that the contour $\Gamma_p$ where $\tilde n^z_k=0$ is a circle centered at $k_c$ which has a radius 
\begin{equation}
k_p= \sqrt{{v_{\textrm{LZ}}\ln 2}/\pi}, \label{eqn:kp}
\end{equation}

We are now in a position to analytically estimate $\tilde\Sigma_{xy}$ in the small $v$ limit. We first rewrite the integration in Eq. (\ref{eqn:A2}) as $\int_{A_2}d^2 k=\int_{\textrm{BZ}}d^2k-\int_{A_1}d^2k$, where the first integration is over the whole BZ while the second over patch $A_1$. Using Eq. (\ref{eqn:lz}), and noting that $\tilde{n}^z_k>0$ in $A_1$ while $\tilde{n}^z_k<0$ in $A_2$, we then find
\begin{equation}
\tilde\Sigma_{A_2}=\frac{1}{2\pi}\int_{\textrm{BZ}}d^2k (1-2e^{-\frac{\pi}{v_{\textrm{LZ}}}k^2})\mathcal F_k^f-\tilde\Sigma_{A_1}.\nonumber
\end{equation}
Therefore, the Hall conductance $\tilde\Sigma_{xy}=\tilde\Sigma_{A_1}+\tilde\Sigma_{A_2}$ becomes
\begin{equation}
\tilde\Sigma_{xy}=\mathcal C-\frac{1}{\pi}\int_{\textrm{BZ}}d^2k e^{-\frac{\pi}{v_{\textrm{LZ}}}k^2}\mathcal F_k^f. \label{eqn:Sigma1}
\end{equation}
Here $\mathcal C=\frac{1}{2\pi}\int_{\textrm{BZ}}d^2k\mathcal F_k^f=1$ is simply the Chern number of the lower band of $H(m_f)$. To evaluate the integral in Eq. (\ref{eqn:Sigma1}), we change to polar coordinates: $(k_x,k_y)\rightarrow (k,\theta)$. In addition, noting that the integrand exponentially decays with $k$, we extend the upper limit of the integration of $k$ to $+\infty$. Equation (\ref{eqn:Sigma1}) then becomes, in terms of new variable $x=k^2$, \begin{equation}
\tilde\Sigma_{xy}=\mathcal C-\frac{1}{2\pi}\int_0^{2\pi} d\theta \int_0^\infty dx e^{-\frac{\pi}{v_{\textrm{LZ}}}x} \tilde{\mathcal F}(x,\theta). \label{eqn:Sigma2}
\end{equation}

Here $\tilde{\mathcal F}(x,\theta)$ denotes the Berry curvature $\tilde F_k^f$ expressed in terms of $(x,\theta)$. In Eq. (\ref{eqn:Sigma2}), the integration over $x$ can be performed through integration by parts, which gives
\begin{eqnarray}
&&\int_0^\infty dx e^{-\frac{\pi}{v_{\textrm{LZ}}}x} \tilde{\mathcal F}(x,\theta)\nonumber\\
&=&\frac{v_{\textrm{LZ}}}{\pi}\tilde F(0,\theta)+\left(\frac{v_{\textrm{LZ}}}{\pi}\right)^2\tilde F'(0,\theta)+\left(\frac{v_{\textrm{LZ}}}{\pi}\right)^3\tilde F''(0,\theta)+...\nonumber
\end{eqnarray}
which forms an expansion in terms of the small parameter $v_{\textrm{LZ}}$.  Here $\tilde{\mathcal F}^{(n)}(0,\theta)$ ($n=0,1...$) is the $n$th derivative of $\tilde{\mathcal F}(x,\theta)$ with respect to $x$ evaluated at $k_c=(0,0)$, which can be readily calculated from Eq. (\ref{eqn:dkc}). Keeping the lowest order in $v_{\textrm{LZ}}=\lvert 2+m_f \rvert v$ in above expansion, we find the leading contribution to Eq. (\ref{eqn:Sigma2}) in terms of $v$ is given by 
\begin{equation}
\tilde\Sigma_{xy}\simeq \mathcal C-\frac{1}{2\pi}\frac{v}{|m_f+2|}+\mathcal O(\frac{v^2}{(m_f+2)^2}). \label{eqn:Sigma5}
\end{equation}

\begin{figure}[h!]
\centering
\includegraphics[width=0.65\columnwidth]{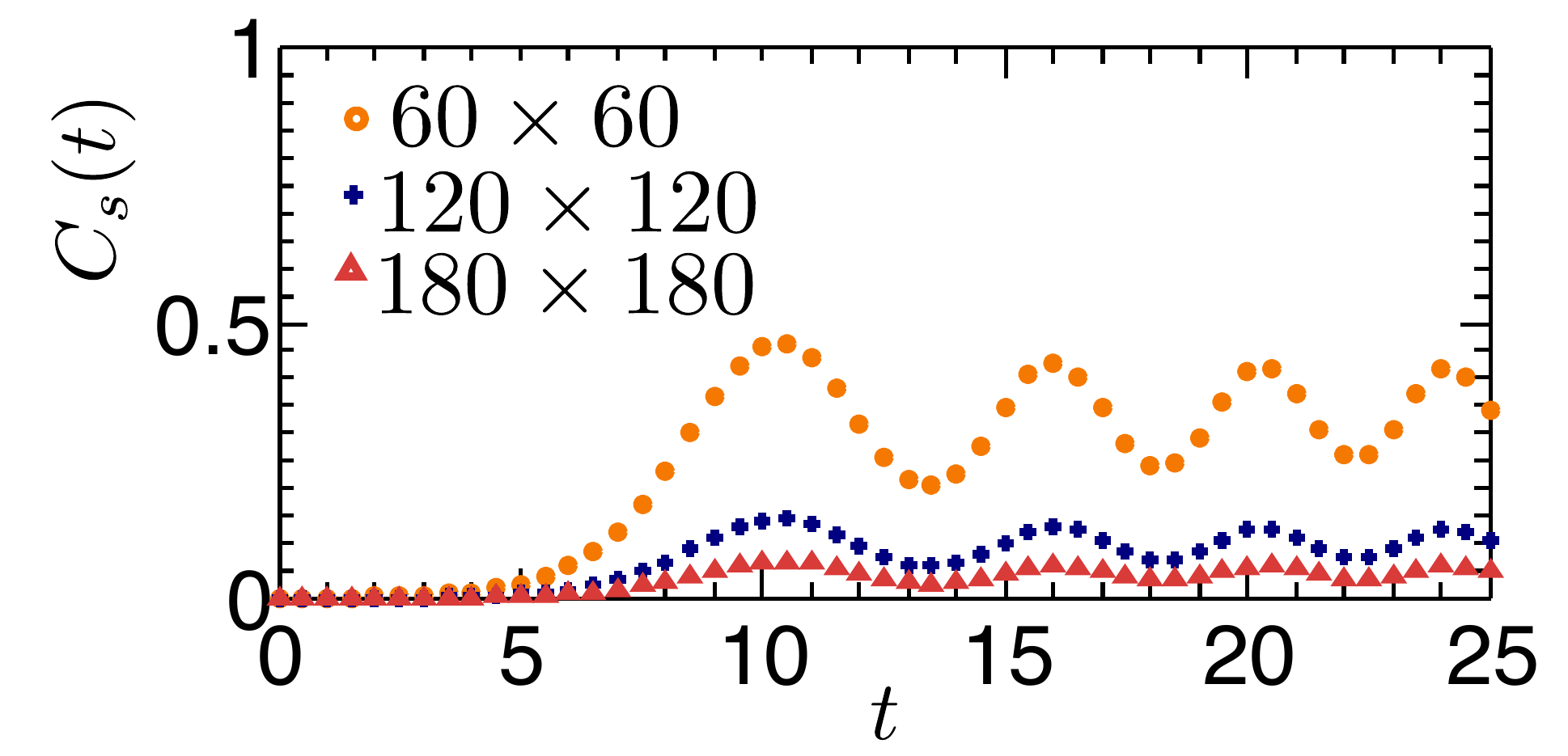}
\caption{\label{fig:one} (color online). Time evolution of Chern number $\mathcal C_s(t)$ for the coherently evolving state simulated for various system sizes.  Right panel: Dephasing dynamics of Hall response $\Sigma_{xy}(t)$ for various system size. In all plots, the system is initialized in the ground state of $H(m_i)$ with $m_i=-2.7$, before $m(t)$ is ramped to $m_f=-1$ as $m(t)=m_i+(m_f-m_i)(1-e^{-0.1t})$. For the right panel, dephasing rate is $\gamma_k=0.15$. }\label{AppFig:2}
\end{figure}

We thus conclude that $\lim_{v\rightarrow 0}\tilde\Sigma_{xy}=\mathcal C$ for a dephased steady state, valid in the thermodynamic limit. In addition, as illustrated in Fig. \ref{AppFig:1}, the linear scaling with $v$ as predicted by Eq. (\ref{eqn:Sigma5})  (see blue curve) is shown to agree well with the numerical results (see orange curve) for small $v$. 

\bigskip

\section{Finite size effects}

In this section, we show how the finite size of the simulated system affect the coherent dynamics of the Chern number $\mathcal C_s(t)$ and the dephasing dynamics of the Hall response. 

In the thermodynamic limit, the Chern number $\mathcal C_s(t)$ of the state is conserved during the coherent evolution. However, simulating the real time dynamics of $\mathcal C_s(t)$ for a system of finite size with periodic boundary conditions, deviations of $\mathcal C_s(t)$ from the constant value occur. To get a feeling for the relevance of finite size effects in our simulations, we compare the time-dependent numerical value of $\mathcal C_s(t)$ when the system Hamiltonian is slowly ramped into topological regime for various system sizes, as shown in Fig. \ref{AppFig:2}. There, we see that $\mathcal C_s(t)$ for a system size of $60\times 60$ sites (see yellow curve) significantly deviates from its zero value in the thermodynamic limit. However, this deviation decreases rapidly with increasing system size. For a system size of $120\times 120$ sites as considered in our simulations on the dephasing dynamics in the main text, the average deviation is as low as $ 0.05$. 

\begin{figure}[h!]
\centering
\includegraphics[width=0.6\columnwidth]{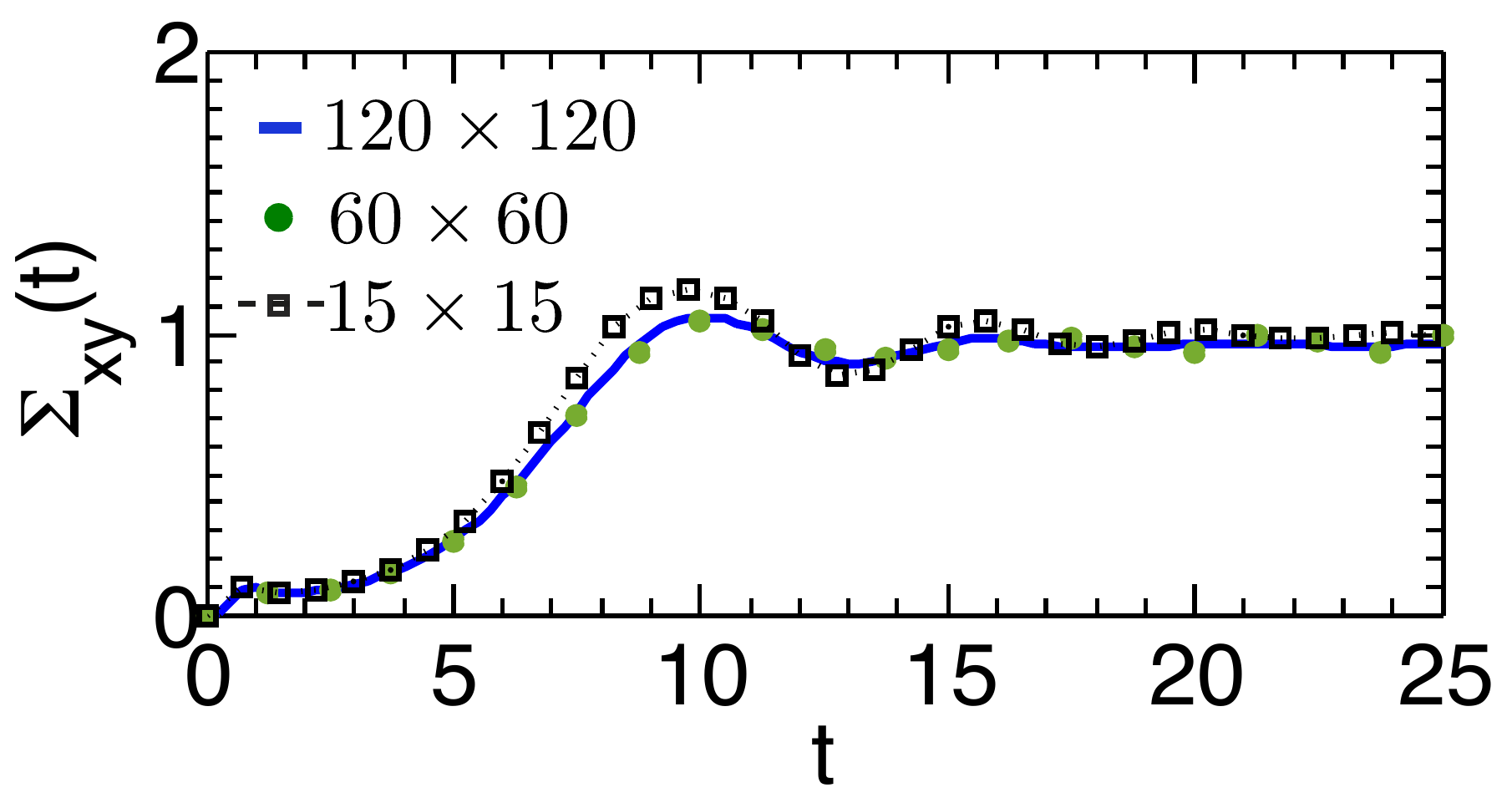}
\caption{\label{fig:one} (color online). Dephasing dynamics of Hall response $\Sigma_{xy}(t)$ for various system sizes. Same ramp protocol as in Fig. \ref{AppFig:2}, dephasing rate is $\gamma_k=0.15$. }\label{AppFig:rampsize}
\end{figure}

Figure \ref{AppFig:rampsize} presents the dephasing dynamics of Hall response for various system size in a slow parameter ramp. There, we see that the Hall response quickly stablizes toward an asymptotically quantized value with increasing system size.   \bigskip

\bigskip
\section{Effects of imperfections}

In this section we present additional data showing that the dephasing dynamics of the Hall repsonse discussed in the main text is robust against $(i)$ weak static disorder and $(ii)$  the presence of an external confining trap, both of which break the translational invariance.  

To this end, we turn to the tight-binding representation of the Dirac Hamiltonian in the main text, which takes the form \cite{Qi2008}
\begin{eqnarray}
H&=&\sum_{n}\Big[c_{n}^\dag \frac{\sigma_z-i\sigma_x}{2}c_{n+\hat{x}}+c_{n}^\dag\frac{\sigma_z-i\sigma_y}{2}c_{n+\hat{y}}+\textrm{H.c.}\Big]\nonumber\\
&+&m(t)\sum_{n\alpha}c_{n}^\dag c_{n}
\end{eqnarray}
where $c_{n}(c^+_{n})$ annihilates (creates) a two-state spinor at site $n=(x_n,y_n)$, and $\hat{x}$ ($\hat{y}$) denotes one lattice spacing along $x$ ($y$) direction.

\begin{figure}[h!]
\centering
\includegraphics[width=\columnwidth]{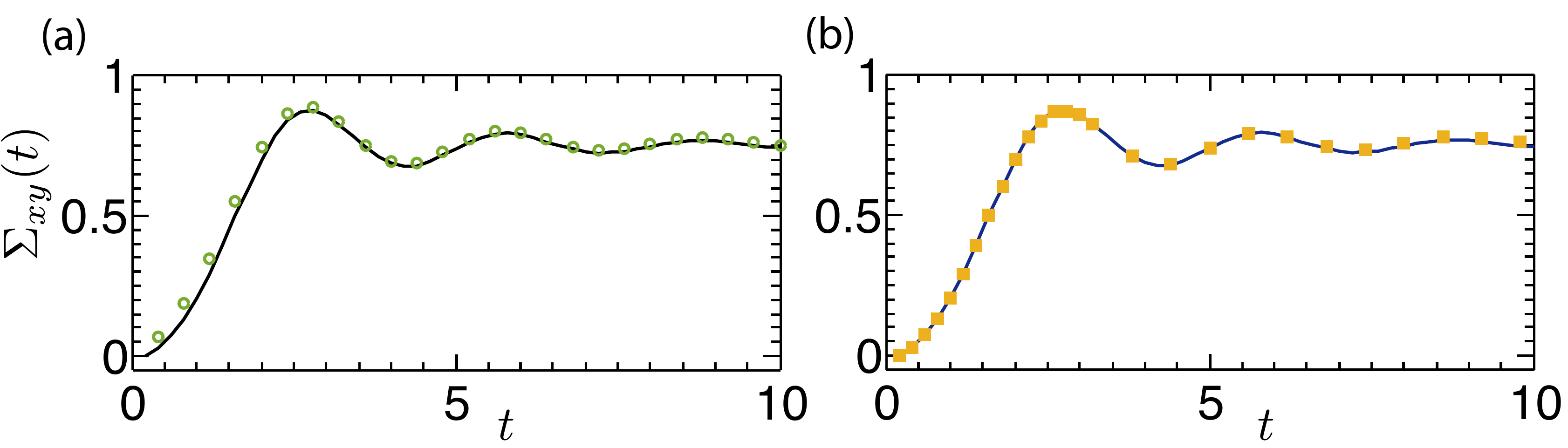}
\caption{\label{fig:one} (color online). Dephasing dynamics of Hall response $\Sigma_{xy}(t)$ (a) with (green circles) and without weak disorder (black curve), (b) with (yellow squares) and without (blue curve) a harmonic trap. In both cases,  the system is initialized in the ground state of $H(m_i)$ with $m_i=-3$, before $m(t)$ is ramped to $m_f=-1$ as $m(t)=m_i+(m_f-m_i)(1-e^{-t})$ with a velocity $v=1$. The dephasing rate is $\gamma=0.2$. For (a), one realization of disordered potential with $W_d=0.02$ (see text) is shown. For (b), we choose a trap potential $V_{tr}=2.8$ (see text). A system of $40\times 40$ is simulated with periodic boundary conditions. }\label{Fig:robust}
\end{figure}

$(i)$ We first consider the effect of weak on-site disorder modelled by $H_{\textrm{dis}}=\sum_{j\alpha}  W_j c_{n\alpha}^\dag c_{n\alpha}$, where $W_j$ is uniformly distributed within  $[-W_d/2,W_d/2]$. In Fig. \ref{Fig:robust} (a), we compare the numerical results for the dephasing dynamics of $\Sigma_{xy}(t)$ with (green circle) and without (dark blue curve) disorder. We see that the Hall response remains stable in the presence of weak disorder. Note that here the deviation of the Hall response from the quantized value is mainly due to the finite ramp velocity $v$ (as can be inferred from Eq. (\ref{eqn:Sigma5})): here we choose $v=1.0$ such that in the corresponding coherent dynamics $\mathcal C(t)$ only deviates from $0$ by approximately $0.06$ due to the finite size effect. 

$(ii)$ Next, we show the influence of an external harmonic trapping potential modelled by $V_{tr}(l_x,l_y)=V_c\left[(x_c-l_x)^2/N_x^2+(y_c-l_y)^2/N_y^2\right]$, with $l_{x(y)}=1,2,..., N_{x(y)}$ and $x_c=(N_x+1)/2$ and $y_c=(N_y+1)/2$ (the lattice constant is set as $a=1$). Typically such confining trap is shallow, so that the potential varies slightly on the period of the lattice. We find that such confinement trap has little effect on the Hall response, as illustrated in Fig. (\ref{Fig:robust}) (b). 

\begin{figure}[h!]
\centering
\includegraphics[width=0.7\columnwidth]{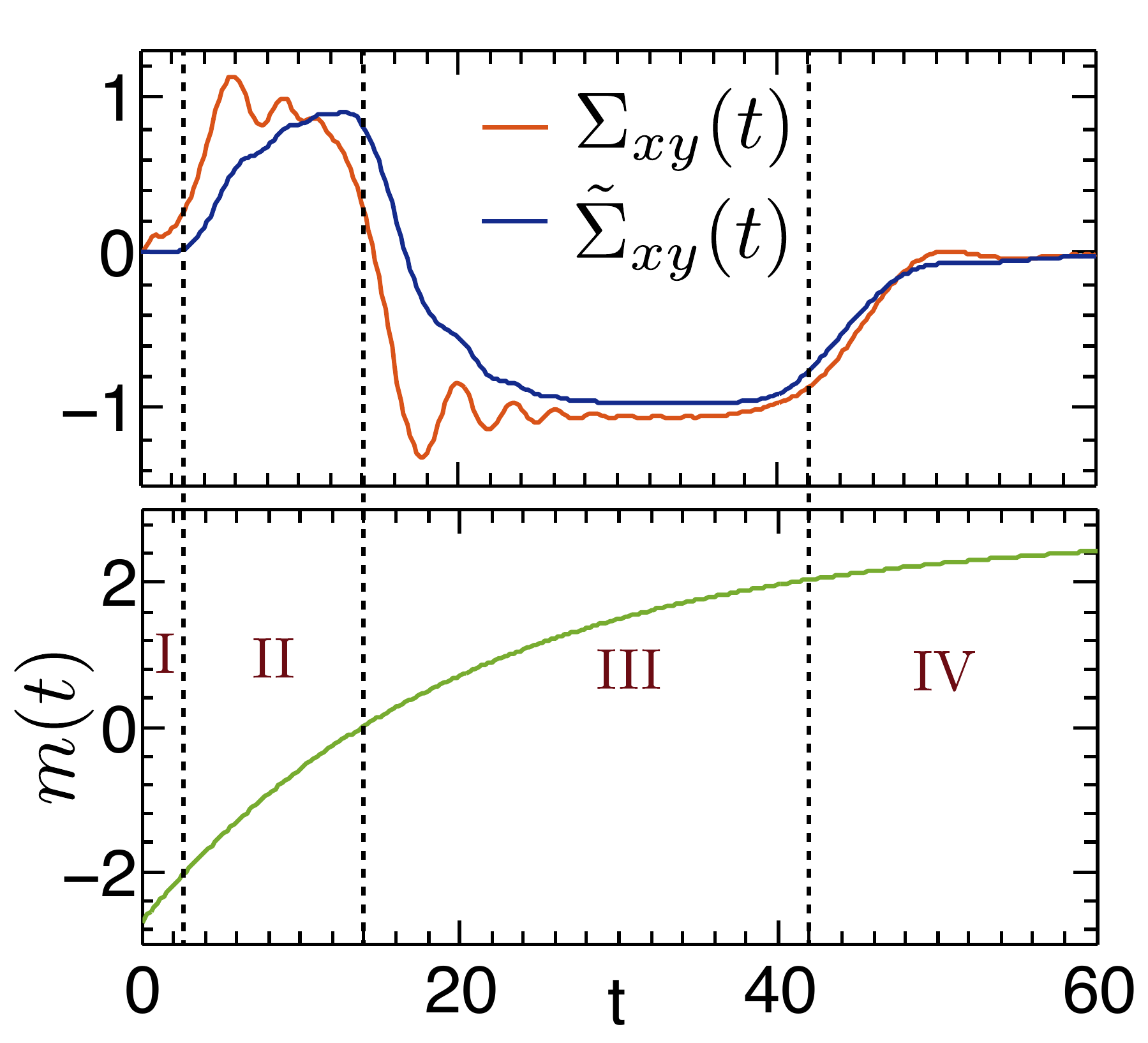}
\caption{(color online). Upper panel: Dephasing dynamics of Hall responses corresponding to a parameter ramp in the lower panel, where $m(t)=m_i+(m_f-m_i)(1-e^{-0.05t})$ with $m_i=-2.7,m_f=2.7$. Lower panel: $m(t)$ traversing for different regions separated by phase transitions. The Chern number of the the system Hamiltonian is $\mathcal C=0$ in region (I), $\mathcal C=1$ in region (II), $\mathcal C=-1$ in region (III), and $\mathcal C=0$ in region (IV). The dephasing rate is $\gamma=0.15$, and system size $120\times 120$.}\label{AppFig:HF}
\end{figure}

\section{Ramp through multiple phase transitions}

Here we provide additional data to further support our claim that the non-equilibrium Hall response uniquely reflects the topology of the instantaneous Hamiltonian. In the upper panel of Fig. \ref{AppFig:HF}, we plot the dephasing dynamics of the Hall response in a parameter ramp shown in the lower panel of Fig. \ref{AppFig:HF}: the mass parameter $m(t)$ is varied in such a way that the Chern number of the Hamiltonian undergoes the sequence $\mathcal C=0\rightarrow 1\rightarrow -1\rightarrow 0$.  We see that the Hall response $\Sigma_{xy}(t)$ in dephasing dynamics  stably follows the topological invariant of the Hamiltonian, a behavior well captured by $\tilde{\Sigma}_{xy}(t)$.

\section{Sudden quench limit and dynamical change of Chern number by active cooling}
\begin{figure}[htp]
\centering
\includegraphics[width=1.02\columnwidth]{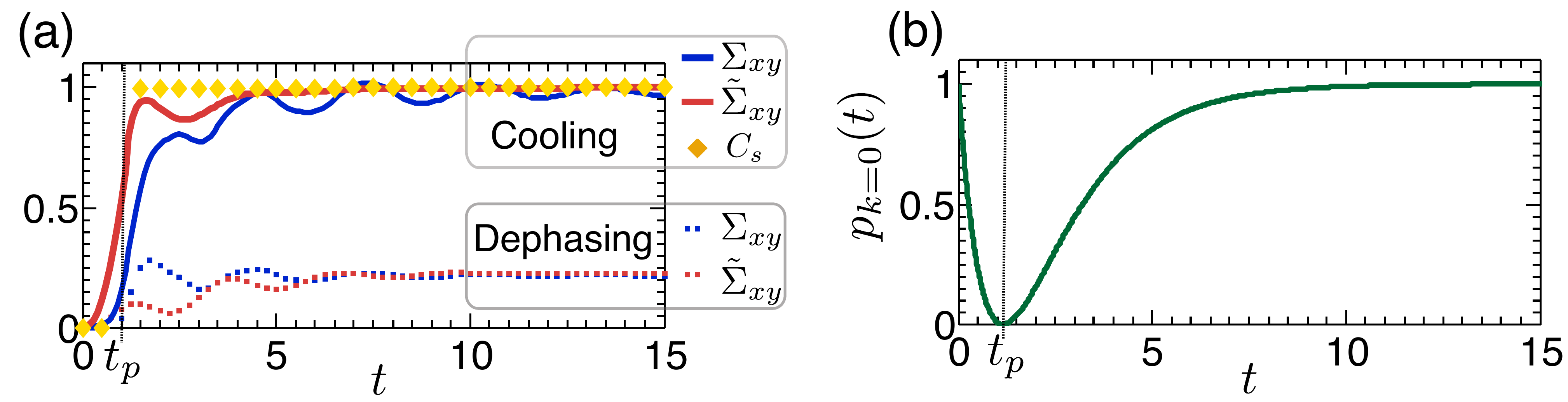}
\caption{\label{fig:four} (color online). Panel (a): Real time dynamics of state Chern number $\mathcal C_s(t)$ and Hall response ($\Sigma_{xy}(t)$, $\tilde \Sigma_{xy}(t)$) of a quenched open system, in the presence of cooling (upper three curves) and dephasing (lower two curves), respectively. Panel (b): Purity gap $p_k(t)$ at $k=(0,0)$ as a function of time with cooling. At $t_p$, where $p_{k=0}(t_p)=0$,  $\mathcal C_s(t)$ jumps from zero to one in (a). Initial state in all plots is the ground state of $H(m_i)$ with $m_i=-4.1$, before the mass parameter is quenched to $m_f=-1$. $\gamma_k=0.25$ for pure dephasing case. $\gamma^c=0.3$ for cooling case [see Eq.~(\ref{eqn:eom})]. System size is $120\times 120$ sites in all plots. }
\end{figure}

\begin{figure}[h!]
\centering
\includegraphics[width=0.65\columnwidth]{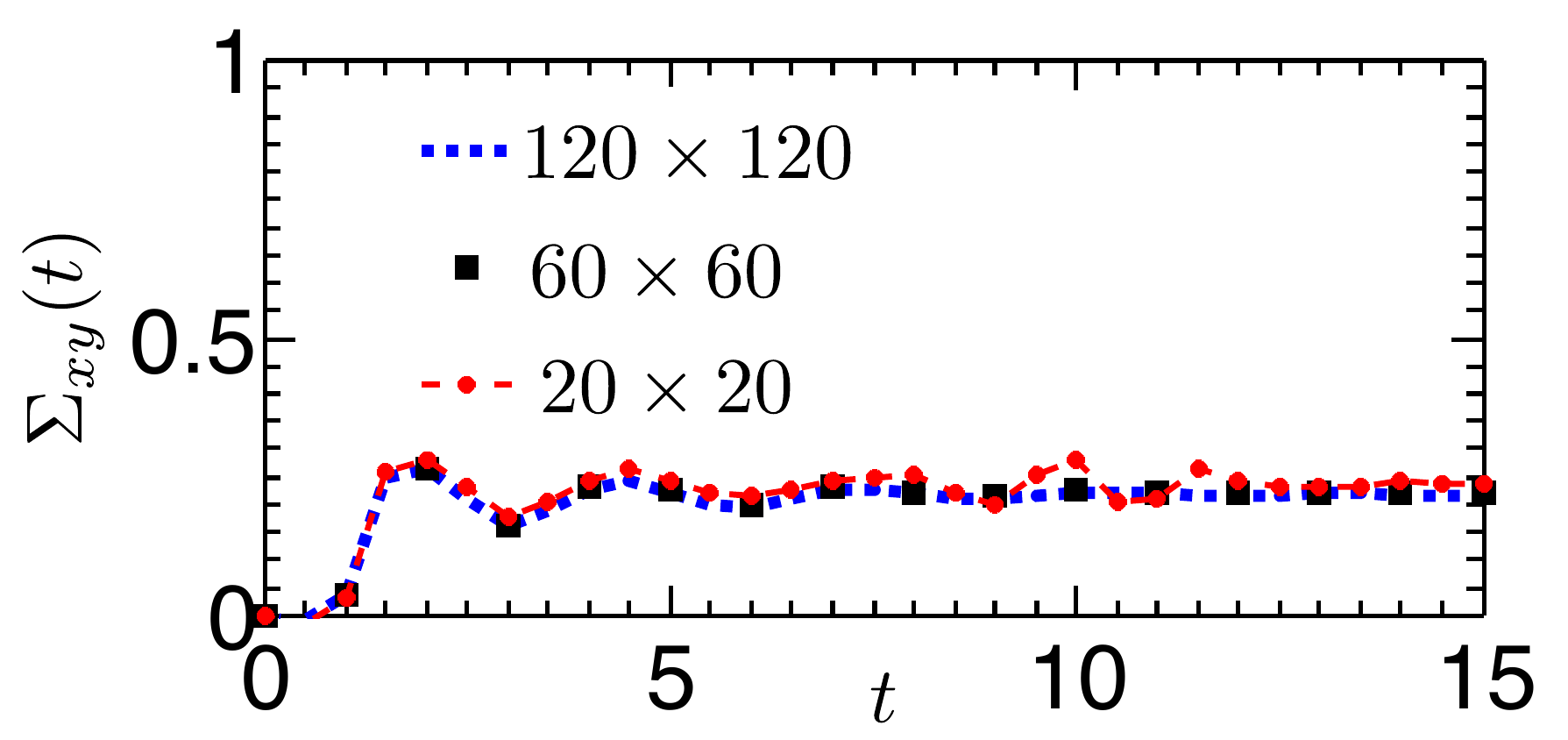}
\caption{\label{fig:one} (color online). Dephasing dynamics of Hall response $\Sigma_{xy}(t)$ for various system size in simulation when Hamiltonian is quenched from $m=-4.1$ to $m=-1$. Dephasing rate $\gamma_k=0.5$.   }\label{AppFig:quenchsize}
\end{figure}

In this section, we present data on the limiting case of a sudden quantum quench ($v\rightarrow \infty$) from a trivial Hamiltonian $H(m_i)$ to a Chern insulator Hamiltonian $H(m_f)$. As is shown in Fig. \ref{fig:four} (a), the Hall response in this case, even in the presence of dephasing, is non-universal as an extensive number of excitations is generated that depends on the specific values of $m_i,m_f$ (The effect of finize size of effect on the dephasing dynamics of Hall response is shown in Fig. \ref{AppFig:quenchsize}). To obtain a quantized Hall response, a cooling mechanism which allows the system to dynamically change its Chern number and to relax entropy into a quantum bath is required. To demonstrate the possibility of such a dynamical transition into a Chern state, we now consider the Lindblad master equation \cite{Lindblad1976,GardinerZoller}
\begin{align}
\dot \rho_k=-i[H(m_f),\rho_k] +\gamma^c\left( 2\tilde \sigma_-\rho_k \tilde \sigma_+-\left\{\tilde \sigma_+\tilde \sigma_-, \rho_k\right\}
\right),
\label{eqn:eom}
\end{align}
where $\tilde \sigma_\pm=(\tilde \sigma_x\pm  i \tilde\sigma_y)/2$ now denote the Pauli matrices in the eigenbasis of the post quench $H(m_f)$. This dynamics allows excitations with respect to $H(m_f)$ to relax into the quantum bath at a rate $\gamma^c$. We expect this to qualitatively capture the essential relaxation processes of a more complicated coupling involving momentum transfer, as microscopically realized by a coupling of atoms in an optical lattice to a surrounding Bose Einstein condensate \cite{Griessner2007} and an electron-phonon coupling, respectively. In Fig.~\ref{fig:four}, we show the dynamical change of the Chern number at the purity gap closing time $t_p$ [see Fig.~\ref{fig:four} (b)] as well as the dynamical buildup of the quantized Hall response [see Fig.~\ref{fig:four} (a)]. Note that the Chern number $\mathcal C_s$ here stays well defined at all times, except $t_p$ where it dynamically changes its value. 

\bibliographystyle{apsrev}